\documentclass[%
 reprint,
 amsmath,amssymb,
 aps,
]{revtex4-2}

\usepackage{graphicx}
\usepackage{dcolumn}
\usepackage{bm}
\usepackage{amsmath}
\usepackage{amssymb}
\usepackage[utf8]{inputenc}
\usepackage[T1]{fontenc}
\usepackage{mathptmx}
\usepackage{xcolor}
\usepackage{enumitem}   
\usepackage{subfig}
\usepackage[format=plain,labelfont=up,textfont=up]{caption}
\usepackage{dsfont}
\usepackage{xr-hyper}
\usepackage{hyperref}
\usepackage{url}
\usepackage{siunitx}
\usepackage{float}
\usepackage{soul,xcolor}

\usepackage{textgreek}

\begin{document}

\title{A Multi-Scale Approach to Simulate the Nonlinear Optical Response of Molecular Nanomaterials}

\author{Benedikt Zerulla*}
\affiliation{Institute of Nanotechnology,
Karlsruhe Institute of Technology (KIT),
76344 Eggenstein-Leopoldshafen, Germany }

\author{Dominik Beutel}
\affiliation{Institute of Theoretical Solid State Physics,
Karlsruhe Institute of Technology (KIT),
76131 Karlsruhe, Germany}
\author{Christof Holzer}
\affiliation{Institute of Theoretical Solid State Physics,
Karlsruhe Institute of Technology (KIT),
76131 Karlsruhe, Germany}

\author{Ivan Fernandez-Corbaton}
\affiliation{Institute of Nanotechnology,
Karlsruhe Institute of Technology (KIT),
76344 Eggenstein-Leopoldshafen, Germany}

\author{Carsten Rockstuhl*}
\affiliation{Institute of Nanotechnology,
Karlsruhe Institute of Technology (KIT),
76344 Eggenstein-Leopoldshafen, Germany}
\affiliation{Institute of Theoretical Solid State Physics,
Karlsruhe Institute of Technology (KIT),
76131 Karlsruhe, Germany}
\author{Marjan Krstić*}
\affiliation{Institute of Theoretical Solid State Physics,
Karlsruhe Institute of Technology (KIT),
76131 Karlsruhe, Germany }

\keywords{nonlinear optics, multi-scale workflow, second-harmonic generation, time-dependent density-functional theory,
transition matrix formalism}

\begin{abstract}
Nonlinear optics is essential for many recent photonic technologies. Here, we introduce a novel multi-scale approach to simulate the nonlinear optical response of molecular nanomaterials combining \textit{ab initio} quantum-chemical and classical Maxwell-scattering computations. In this approach, the first hyperpolarizability tensor is computed with time-dependent density-functional theory and translated into a multi-scattering formalism that considers the optical interaction between neighboring molecules. A novel object is introduce to perform this transition from quantum-chemistry to classical scattering theory: the Hyper-Transition(T)-matrix. With this object at hand, the nonlinear optical response from single molecules and also from entire photonic devices can be computed, incorporating the full tensorial and dispersive nature of the optical response of the molecules. To demonstrate the applicability of our novel approach, the generation of a second-harmonic signal from a thin film of an Urea molecular crystal is computed and compared to more traditional simulations. Furthermore, an optical cavity is designed, which enhances the second-harmonic response of the molecular film by more than two orders of magnitude. Our approach is highly versatile and accurate and can be the working horse for the future exploration of nonlinear photonic molecular materials in structured photonic environments.
\end{abstract}

\maketitle

\section{Introduction and summary}
Nonlinear optics is vital in a large range of today's photonic technologies, and the necessary materials are a key aspect of these developments. Applications of nonlinear optics range from solitons and Kerr frequency comb generation for high-speed data transmission using materials such as MgF$_2$ and Si$_3$N$_4$ \cite{iet:/content/journals/10.1049/el_19910796,Herr:14,Marin:17}, over electro-optic modulators based on materials such as Lithium Niobate \cite{826874}, optoelectronic signal processing in wireless THz networks \cite{Harter:19}, and high-precision frequency metrology \cite{Udem:02,doi:10.1126/science.aay3676} to direct-write 3D laser lithography \cite{micro1020013}. Also, quantum optics information processing and second-harmonic efficiency enhancement that exploit nanostructured nonlinear materials are of interest \cite{Liu:18,doi:10.1021/acsphotonics.2c00835,doi:10.1021/acsphotonics.7b01478}. A strong second-harmonic generation (SHG) response is also shown by nanocrystalline silicon nanoparticles \cite{doi:10.1021/acs.nanolett.7b00392}. In \cite{Rocco:21}, furthermore, a metasurface is presented that controls SHG emission via thermal effects.

Already from the beginning of the study of nonlinear optical phenomena, molecular materials have been of particular interest, and they continue to be crucial \cite{Gu:16}. The reason for that is the sheer unprecedented opportunity, thanks to the large chemical design space, to tailor molecular materials and to lend them with properties on demand. The linear and nonlinear optical molecular properties are at stake. These properties should be controlled qualitatively. This is understood here as the symmetries dictating which entries of a tensor, that represents the anisotropic properties, are non-zero. On the other hand, the properties should be controlled quantitatively, i.e., how large the non-zero coefficients are. Nonlinear molecular materials are used for the multiphoton absorption \cite{doi:10.1021/cr050054x}. Also, organic chromophore-containing polymers whose refractive index can be modified by light find use \cite{Mar:97}. Other nonlinear molecular materials are surface-anchored metal-organic frameworks for the second-harmonic generation \cite{https://doi.org/10.1002/adma.202103287} or metal-assisted chromophores showing two-photon absorption \cite{doi:10.1021/ct500579n}.

Information on the actual material parameters required to describe nonlinear optical phenomena in a qualitative and quantitative sense is scarce. In most cases, designing devices that exploit nonlinear materials and explaining experimental observations can reliably be done by solving Maxwell's equations equipped with nonlinear constitutive relations. Here, the materials are described at the macroscopic level using a nonlinear susceptibility \cite{doi:10.1021/acs.nanolett.7b01488}. Sometimes, it suffices to estimate the nonlinear susceptibility on phenomenological grounds. Alternatively, it can also be measured and retrieved experimentally \cite{Boudebs:01,Wolf:18,BOYD201474}. However, here, we encounter limitations. The full tensorial nature is hard to assess, and in most cases, the properties are retrieved only at some discrete wavelengths, and their dispersion remains elusive. Therefore, computational approaches are required to make the nonlinear optical material properties available. While for crystalline solid-state materials that information is available by now, a comparable development for molecular materials is much more in its infancy. Especially when it comes to nonlinear molecular materials, a reliable description is urgently needed since the vast design space does not permit a systematic experimental characterization in a reasonable amount of time. Instead, explorations {\it in silico} are urgently needed to provide (a) reliable and complete quantitative and qualitative information on the nonlinear properties, (b) to systematically assess and compare different possible molecular materials, and (c) to ultimately enable the accurate study of functional photonic devices made from these nonlinear photonic molecular materials. 

To predict the nonlinear response of a nanomaterial of a macroscopic molecular structure, a multi-scale ansatz is required. The point of departure for such a multi-scale ansatz are quantum-chemical simulations that capture the properties of the molecules. This microscopic molecular information must then be connected to the macroscopic nonlinear optical response of the device made from these nonlinear molecular materials. 

In this contribution, we introduce such a multi-scale ansatz. Our ansatz starts with calculating the hyperpolarizability of an individual molecule, which we do using time-dependent density functional theory. Then, the outcome from these quantum chemical simulations needs to be converted into a representation that can flexibly be used and is versatile in the optical modeling part. When considering the linear response, the transition (T-) matrix of a localized object, typically called a scatterer, is a comprehensive representation of its optical response to an incident electromagnetic wave \cite{Waterman1965}. For molecules, their T-matrix can be computed from its polarizability tensor \cite{Fernandez-Corbaton:2020}. Based on this T-matrix, the linear response of molecular structures in photonic devices can be obtained \cite{SURMOFCavity,https://doi.org/10.1002/adfm.202301093}. 

In here, we introduce a novel object called the Hyper-T-matrix to express the nonlinear optical response of an individual molecule. The Hyper-T-matrix can be computed from the hyperpolarizability of a molecular scatterer. We then continue to describe the nonlinear optical response of a molecular crystalline material. The molecules continue to be described at an individual level using the Hyper-T-matrix, and the optical response from the entire materials is described in a multiple scattering framework accounting for the crystalline lattice. 

Compared to more traditional approaches that exploit effective medium properties, there are several advantages of our presented method that continues to treat each molecule as a nonlinear optical scatterer. First, the scattering properties of the nonlinear molecule used to calculate the response of the macroscopic structure are computed \textit{ab initio}. We concentrate here on second-order nonlinear properties, but higher-order nonlinearities can be considered as well. The nonlinear response is not estimated on phenomenological grounds, and approximations used to calculate effective quantities such as the nonlinear susceptibility, e.g., that neglect the optical interactions between molecules forming the material, are neither necessary nor applied. Analogously to the multi-scale ansatz for the linear response of molecular structures, these aspects render the presented approach very useful to compare the simulated nonlinear response of the macroscopic structure to experimentally measured spectra. We emphasize that while currently the nonlinear optical response at only a few wavelengths can be experimentally studied, due to the constraints on available lasers for nonlinear optics, and the full tensorial details of the nonlinear response are barely retrieved, our multi-scale ansatz enables the theoretical analysis of a spectral range, which might be addressable by future laser sources, and it offers full tensorial information. Moreover, our approach applies to cases where the description of the material on the base of a bulk susceptibility is questionable, e.g., in molecular monolayers.

The article is structured as follows. Section~\ref{sec:HyperTMatrix} introduces the Hyper-T-matrix, its computation from the hyperpolarizability, and how it is used to calculate the SHG response of layers of periodically arranged molecules and stacks thereof, which would form an entire material. Section~\ref{sec:SHGUrea} exemplifies the application of our computational framework. We study the nonlinear optical response of a prototypical second-order nonlinear molecular material, Urea, and compare predictions to the response calculated with a finite-element method-based solver that considers the material at the level of an effective medium. Furthermore, as an example of an advanced photonic device, we study the SHG response of a film of the Urea molecular crystal placed inside a planar metallic cavity. Here, the SHG signal is considerably enhanced due to simultaneous cavity resonances at the fundamental and the SHG frequency. Section~\ref{sec:Conclusion} concludes our work and provides an outlook.

\section{Second harmonic response of a molecular nanomaterial based on its Hyper-T-matrix}\label{sec:HyperTMatrix}
The T-matrix of a scattering object relates the multipolar field coefficients of an incident field to the coefficients of a scattered field \cite{Waterman1965}. The fundamentals of the T-matrix formalism are explained in the first section of the SI. In a nutshell, it is a versatile framework to predict the optical response from photonic nanomaterials made from discrete objects. In \cite{SEKULIC2021107643,Sekulic2022}, the T-matrix formalism has also been extended to calculate the second-harmonic generation (SHG) of finite clusters of spheres and finite clusters of arbitrarily shaped objects made from materials characterized by a second-order nonlinearity. In these cases, the material properties were described using a second-order susceptibility. 

In this article, we present how to compute the nonlinear optical response of a macroscopic device containing molecular crystals based on \textit{ab initio} quantum-chemical methods. In this context, the quantum-chemical results are directly used to compute the linear and nonlinear scattering response of the individual molecules. The nonlinear optical response from an entire device can be calculated using the general T-matrix formalism. We stress that the medium will not be described by a macroscopic bulk nonlinear susceptibility that, except from indirect measurements, can only be computed with approximations or even estimated with phenomenological approaches.

In \cite{Fernandez-Corbaton:2020}, the relation between the polarizability tensor calculated with quantum-chemical methods and the T-matrix is derived to compute the linear optical response of molecules. In the following, we derive analogously the relation between the hyperpolarizability tensor and a new quantity, which we call the Hyper-T-matrix. It relates the multipolar expansion coefficients of two incident electric fields at the fundamental frequency to the multipolar expansion coefficients of the scattered electric field at the SHG frequency. 

The first or quadratic hyperpolarizability tensor is an object describing the induced dipole moment of a molecule due to an external electric field based on second-order nonlinear effects, see \cite{PhysRevA.26.2028,Singer:87,C8TC05268A,https://doi.org/10.1002/adma.202103287} for instance. For SHG processes, two photons of one or two beams with electric fields $\bm{E}_{1}(\omega)$ and $\bm{E}_{2}(\omega)$ and  with the same fundamental frequency $\omega$, generate a single photon with twice the frequency $\Omega=2\omega$ upon interaction with the molecule. A component of the induced electric dipole moment $\bm{p}$ of the molecule reads
\begin{align}\label{eq:SHGDipoleMoment}
    p_i(\Omega) = \frac{1}{2}\sum_{j,k}\beta_{ijk}(-\Omega;\omega,\omega)E_{1j}(\omega)E_{2k}(\omega)\mathrm{.}
\end{align}
Here, $\mathbf{\beta}(-\Omega;\omega,\omega)$ is the first hyperpolarizability in the Cartesian basis describing the SHG process. Even though we consider in this article for simplicity the SHG process, one can extend the approach to other second-order nonlinear processes or to higher nonlinear processes such as third-order processes. That would require knowledge of the cubic hyperpolarizability $\gamma_{ijkl}$ \cite{Singer:87,C8TC05268A}. Further, note that we concentrate on processes where only electric fields are important and do not induce any magnetic dipole moments. 

Equivalently to Equation~(S11) from \cite{Fernandez-Corbaton:2020}, we relate the dipolar expansion coefficients $\bm{c}^{\Omega}$ of the scattered electric field at the SHG frequency $\Omega$ to the dipole moment of the scatterer via
\begin{align}\label{eq:RelScatDipol}
    \begin{pmatrix}
    c_{1-1}^{\Omega}\\
    c_{10}^{\Omega}\\
    c_{11}^{\Omega}
    \end{pmatrix}
    =\frac{c_{\mathrm{h}}(\Omega)Z_{\mathrm{h}}(\Omega)(k_{\mathrm{h}}\left(\Omega\right))^3}{\sqrt{6\pi}}\bm{p}(\Omega)\mathrm{,}
\end{align}
where $c_{\mathrm{h}}(\Omega)=1/\sqrt{\varepsilon_{\mathrm{h}}(\Omega)\mu_{\mathrm{h}}(\Omega)}$ is the speed of light in the surrounding achiral medium, \newline $Z_{\mathrm{h}}(\Omega)=\sqrt{\mu_{\mathrm{h}}(\Omega)/\varepsilon_{\mathrm{h}}(\Omega)}$ its wave impedance, and $k_{\mathrm{h}}(\Omega)=\Omega\sqrt{\varepsilon_{\mathrm{h}}(\Omega)\mu_{\mathrm{h}}(\Omega)}$ its wave number. $\epsilon_{\mathrm{h}}(\Omega)$ is the permittivity and $\mu_{\mathrm{h}}(\Omega)$ is the permeability of the surrounding host medium.

From Equation~(\ref{eq:SHGDipoleMoment}), the relation between the electric dipole moment and the hyperpolarizability is known. We now define the Hyper-T-Matrix $\mathbf{T}^{\mathrm{Hyper}}(-\Omega;\omega,\omega)$ in the spherical vector basis such that
\begin{align}\label{eq:RelScatHyper}
    c_{1m}^{\Omega}
    = \sum_{r,s} \mathrm{T}^{\mathrm{Hyper}}_{mrs}(-\Omega;\omega,\omega)a_{1r}(\omega)a_{2s}(\omega)\mathrm{,}
\end{align}
where $a_{1r}(\omega)$ and $a_{2s}(\omega)$ are the multipolar expansion coefficients of the incident electric fields $\bm{E}_{1}(\omega)$ and $\bm{E}_{2}(\omega)$. Note that the Hyper-T-Matrix $\mathbf{T}^{\mathrm{Hyper}}(-\Omega;\omega,\omega)$ is a third-rank tensor, like the first hyperpolarizability and the second-order nonlinear susceptibility.

With Equation~(S9) from \cite{Fernandez-Corbaton:2020}, one derives
\begin{align}\label{eq:SphericalFieldCartField}
    a_{1,2;r,s}(\omega) = -\mathrm{i}\sqrt{\frac{12\pi}{2}}\bm{\hat{e}}_{r,s}^{\dagger}\cdot\bm{E}_{1,2}(\omega)\mathrm{,}
\end{align}
where $\bm{\hat{e}}_{r,s}$ is a spherical basis vector and $\dagger$ denotes complex conjugation and transposition. $r,s$ are angular momentum indices. Using Equation~(\ref{eq:SphericalFieldCartField}), one can replace the multipolar expansion coefficients $a_{1,2;r,s}$ in Equation~(\ref{eq:RelScatHyper}) by the electric fields of the incident beams. Inserting the result into $c_{1m}^{\Omega}$ on the left-hand side of Equation (\ref{eq:RelScatDipol}), and replacing $\bm{p}(\Omega)$ on the right-hand side of Equation (\ref{eq:RelScatDipol}) by the expression given in Equation~(\ref{eq:SHGDipoleMoment}), one can calculate the Hyper-T-matrix from the hyperpolarizability. The determining equation reads
\begin{align}\label{eq:HyperTHyperPol}
\begin{split}
    &-6\pi\mathbf{C}^{-1}\sum_{m,r,s}\bm{\hat{e}}_{m}\mathrm{T}^{\mathrm{Hyper}}_{mrs}(-\Omega;\omega,\omega)\bm{\hat{e}}_{r}^{\dagger}\cdot \bm{E}_{1}(\omega)\bm{\hat{e}}_{s}^{\dagger}\cdot \bm{E}_{2}(\omega) \\ &= \frac{c_{\mathrm{h}}(\Omega)Z_{\mathrm{h}}(\Omega)(k_{\mathrm{h}}\left(\Omega\right))^3}{2\sqrt{6\pi}}\sum_{i,j,k}\bm{\hat{e}}_{i}\beta_{ijk}(-\Omega;\omega,\omega)E_{1j}(\omega)E_{2k}(\omega),
    \end{split}
\end{align}
where $\bm{\hat{e}}_{i}$ is a unit vector in the Cartesian basis. $\mathbf{C}^{-1}$ is a matrix transforming the vector from the spherical to the Cartesian basis \cite{Fernandez-Corbaton:2020}. Equation~(\ref{eq:HyperTHyperPol}) is solved for the Hyper-T-matrix in the spherical basis, knowing the first hyperpolarizability in the Cartesian basis. Note that while solving Equation~(\ref{eq:HyperTHyperPol}), the electric fields cancel out so that the Hyper-T-matrix does not depend on the incident fields.

With the Hyper-T-Matrix of a molecule at hand, the nonlinear optical response of more complex molecular materials can be studied. Besides investigating the response from individual molecules, we could study the response from an ensemble of molecules by solving the multiple-scattering problem. Moreover, crystalline molecular materials can be easily treated. For example, reflection and transmission from a crystalline layer, i.e., a periodic arrangement of molecules within one plane, can be explored. In such a formalism, substrate, cladding, and conventional optical layers, either homogeneous or nanostructured, can be considered. Also, stacking individual layers allows us to describe layers of a bulk crystalline material. For linear optical effects such as the absorption and circular dichroism of metallic cavities filled with molecular films or metasurfaces made from molecular structures, we have presented simulations and experimental results in \cite{SURMOFCavity,https://doi.org/10.1002/adfm.202301093}.

Here, we strive to describe the response of crystalline layers of molecular materials integrated into a stratified architecture as an example. In such a situation, first, the linear optical problems at the incident fundamental electric fields $\bm{E}_{1}(\omega)$ and $\bm{E}_{2}(\omega)$ are solved with the in-house developed code mpGMM \cite{Beutel:21}. An extended program suite for linear T-matrix calculations is treams \cite{treams}. Solving the linear optical problems, for every two-dimensional lattice of molecules the fundamental electric fields incident on a scatterer in that specific lattice resulting from the fundamental scattering are computed. In the second step, those fields are used to compute the nonlinear optical scattering from one unit cell with Equation~(\ref{eq:RelScatHyper}). In this regard, we use the undepleted pump approximation, which assumes that the fundamental wave is not influenced by the SHG processes.

With Equation~(4) from the Supporting Information, the multipolar coefficients of the SHG scattered electric field of a molecular lattice
\begin{align}
    \bm{c}^{\Omega}_{\mathrm{tot}} = \left(\mathds{1}-\mathbf{T}(\Omega,\Omega)\sum_{\bm{R}\neq 0}\mathbf{C}^{(3)}(-\bm{R})\mathrm{e}^{\mathrm{i}\bm{k}_{\parallel}(\Omega)\bm{R}}\right)^{-1}\bm{c}^{\Omega}
\end{align}
are computed. $\mathbf{C}^{(3)}(-\bm{R})$ is a matrix of translation coefficients of vector spherical waves. $\mathbf{T}(\Omega,\Omega)$ is the (linear) T-matrix at the SHG frequency. $\bm{k}_{\parallel}(\Omega)$ is the component of the wave vector of the zeroth diffraction order of the scattered SHG field parallel to the two-dimensional lattice. Equation~(21) from \cite{Beutel:21} gives the SHG scattered electric field of the two-dimensional lattice in a plane wave decomposition. Using the Q-matrices defined in Equations~(6)-(9) and (12a,b) of \cite{Beutel:21} for the fields at the SHG frequency, the SHG fields leaving in the direction of reflection and transmission of a combined stack of molecular lattices and isotropic linear optical slabs can be computed. Such a combined stack can be a metallic cavity with a molecular filling medium showing a nonlinear optical response, for instance. Additionally, the fields originating from every individual lattice can be calculated, which gives further insights into the optical properties of the macroscopic molecular structure.

The next section demonstrates the usability of the approach by computing the SHG response of a Urea molecular crystal in free space and when placed inside a metallic cavity.

\section{SHG signal of the Urea molecular crystal}\label{sec:SHGUrea}
\begin{figure*}
\centering
	\includegraphics[width=0.8\textwidth]{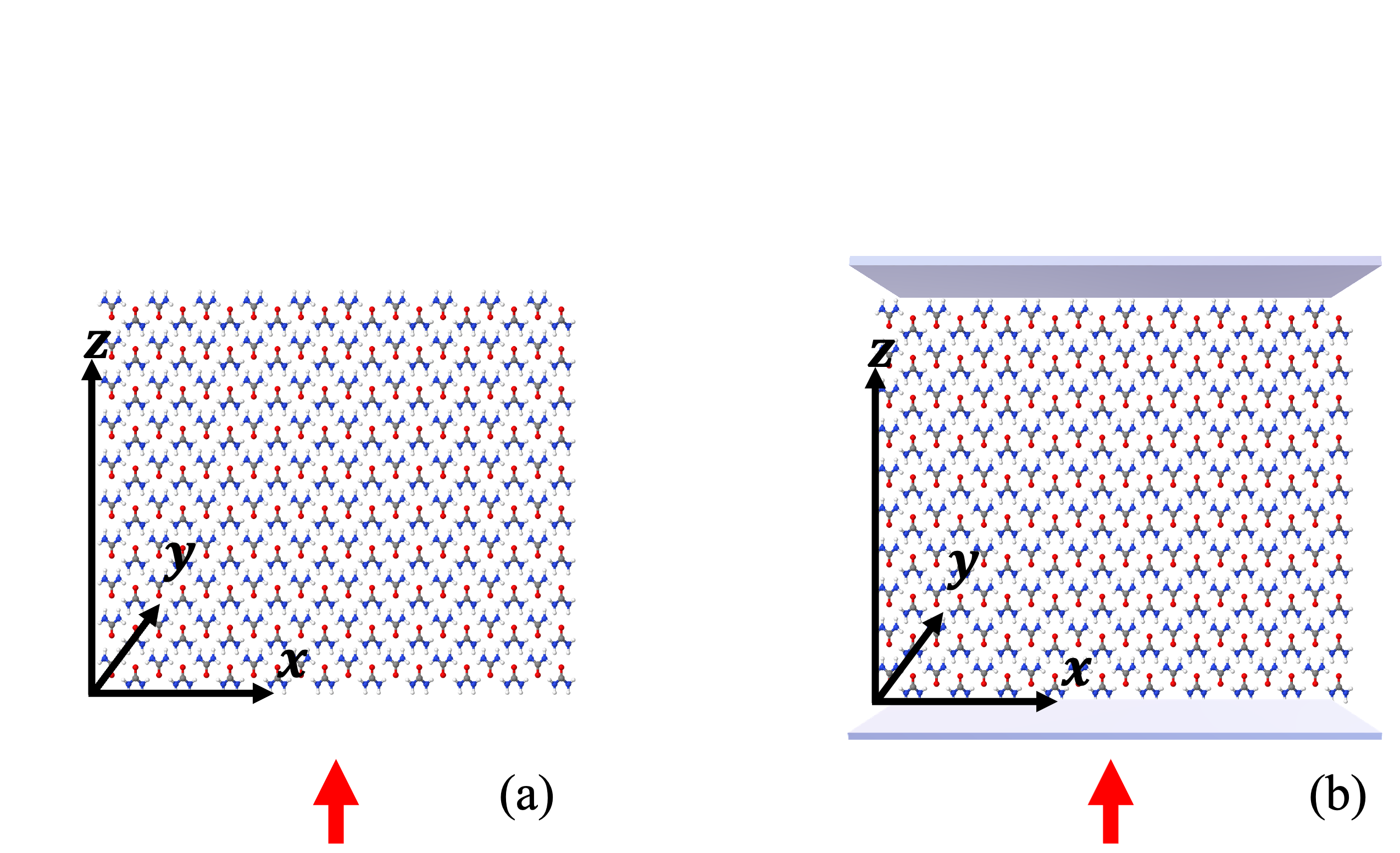}

	\caption{\textbf{(a)} Thin film of the Urea molecular crystal. The thin film is illuminated from below with plane waves of different polarizations and angles of incidence. In the chosen coordinate system, the incident field propagates into the positive z-direction.  \textbf{(b)} Cavity of silver mirrors separated by a thin film of the Urea molecular crystal. The lower mirror has a thickness of 10\,nm, the upper mirror has a thickness of 30\,nm. The cavity is equally illuminated from below with a plane wave linearly polarized in x-direction under normal incidence. The amplitudes of the emanating SHG plane waves propagating into the positive and negative z-direction for both setups are simulated. The images are not to scale, and a reduced number of molecular layers is shown.}
    \label{fig:Setup}
	\end{figure*}

In this section, we show the SHG response of a Urea molecular crystal both in free space and in an optical cavity. The setup analyzed in both situations is depicted in Figure~\ref{fig:Setup}.
For the Urea crystalline film in free space, we compare the results to computations performed with the finite-element based program suite COMSOL Multiphysics \cite{Comsol}. In these finite-element simulations, the molecular film is modeled by linear effective material parameters calculated with the homogenization approach from \cite{https://doi.org/10.1002/adom.202201564} and by a nonlinear effective susceptibility. The effective second-order susceptibility calculated according to Equation~(1.1) from \cite{PhysRevA.26.2028} and Equation~(5) from \cite{C8TC05268A}
\begin{align}\label{eq:chi}
\chi^{(2)}_{ijk}(-\Omega;\omega,\omega) = \frac{1}{\varepsilon_0 V}f_i(\Omega)f_j(\omega)f_k(\omega)\beta_{ijk}(-\Omega;\omega,\omega),
\end{align}
assumes that $\mathbf{\beta}(-\Omega;\omega,\omega)$ is the first hyperpolarizability of the crystalline unit cell and that its axes are the same as the one of the crystal. $V$ is the volume of the unit cell. $f_i(\Omega)$, $f_j(\omega)$, and $f_k(\omega)$ are local field factors given by Equation~(6) from \cite{C8TC05268A},
\begin{align}
    f_i(\omega) = \frac{\left(n_{ii}(\omega)\right)^2+2}{3},
\end{align}
where $n_{ii}(\omega)$ is the refractive index for a molecular crystal axis $i$. In \cite{C8TC05268A}, it is calculated with Lorentz-Lorenz theory. We compute it based on the effective permittivity calculated with the homogenization approach presented in \cite{https://doi.org/10.1002/adom.202201564}, which considers the interaction between different molecules forming the lattice. The effective relative permeability is close to one across the entire frequency range, i.e., that dispersion can be neglected. Note that for the computation of $\mathbf{\chi}(-\Omega;\omega,\omega)$, similar to \cite{C8TC05268A}, we neglect the imaginary part of the refractive index as the absorption of Urea is tiny in this frequency range. In Equation~(\ref{eq:chi}), the interaction between different unit cells in the nonlinear regime is neglected. The nonlinear polarization resulting from this second-order susceptibility is given by
\begin{align}\label{eq:Polarization}
P_i(\Omega) = \frac{1}{2}\varepsilon_0\sum_{j,k}\chi_{ijk}^{(2)}(-\Omega;\omega,\omega)E_{1j}(\omega)E_{2k}(\omega)\mathrm{,}
\end{align}
see Equation~(4) from \cite{C8TC05268A}, for instance. Equation~(\ref{eq:Polarization}) can be directly considered in COMSOL to compute the SHG response of a film consisting of an effective medium with a specific thickness.

For the computation of the Hyper-T-matrix and the effective second-order susceptibility, the first hyperpolarizability is required. 

\subsection{DFT calculations of the first hyperpolarizabilities of Urea}
We follow the same workflow as described in \cite{SURMOFCavity}, except that in addition to the polarizability, the first electric-electric hyperpolarizability $\mathbf{\beta}$ is computed with time-dependent density functional theory (TD-DFT) and used to obtain the Hyper-T-matrix. From the Hyper-T-matrix, the nonlinear response of a photonic device is calculated. Out of the first hyperpolarizability, the Hyper-T-matrix is computed as described above. 

\begin{figure*}
\centering
	\includegraphics[width=1.0\textwidth,keepaspectratio]{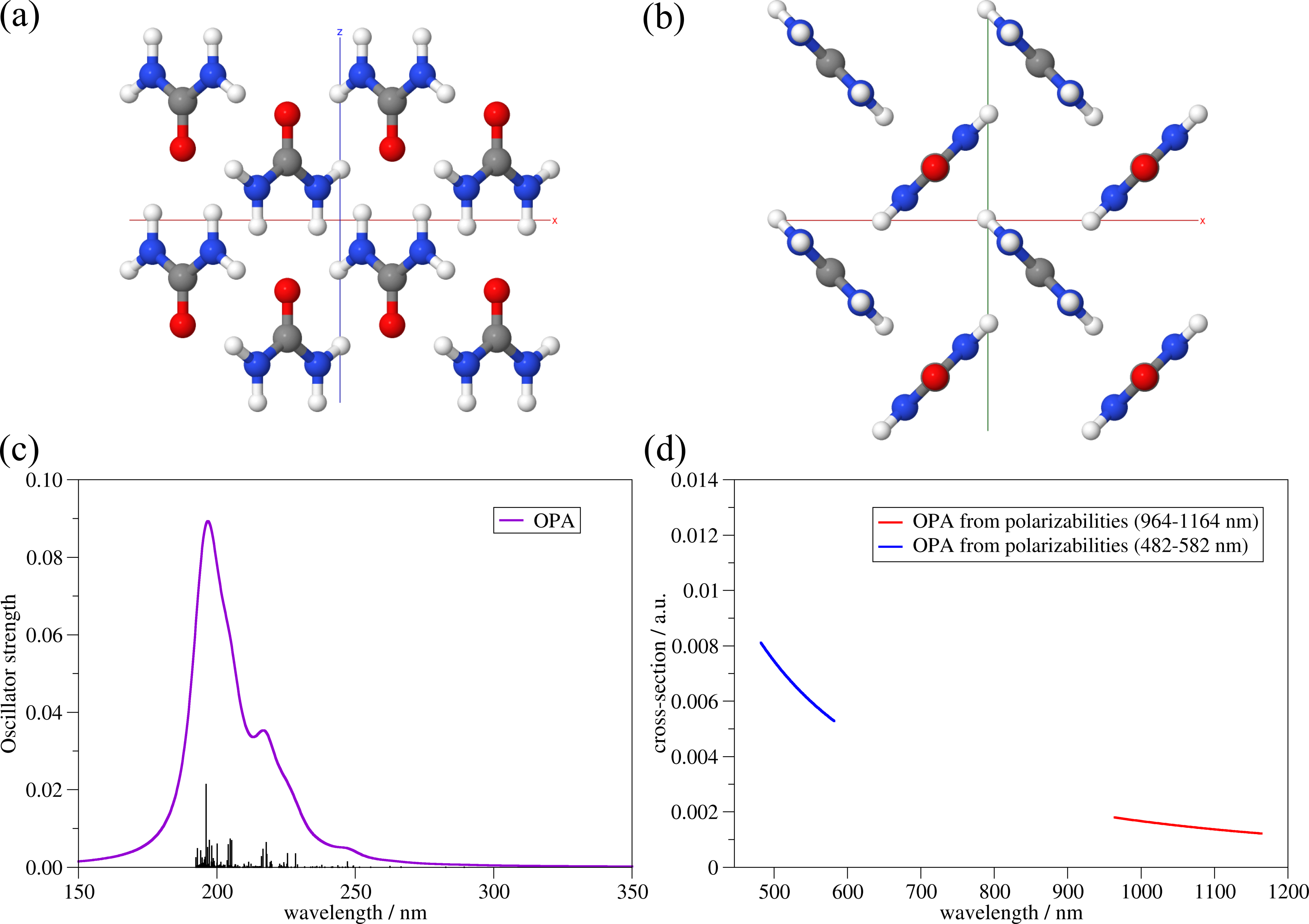}
 \hspace{0cm}
 
	\caption{\textbf{(a)} Side-view of the finite-size molecular model of crystalline Urea used in TD-DFT calculations. White, gray, blue, and red spheres represent hydrogen, carbon, nitrogen and oxygen atoms, respectively.  \textbf{(b)} Top-view of the same model. \textbf{(c)} One-photon absorption (OPA) spectrum of the molecular model calculated from broadened discrete electronic transitions (violet). \textbf{(d)} The OPA cross-section in atomic units is calculated from dynamic polarizabilities for the selected spectral ranges. The shape represents the tail of the Lorentzian-shape broadening function of high-energy electronic transitions below 300 nm.}
    \label{fig:Urea_OPA}
	\end{figure*}

We start by defining the finite-size atomistic model of the crystalline Urea material. Here, we consider the T-matrix and Hyper-T-matrix of a unit cell of the molecular crystal consisting of a larger number of molecules rather than the T-matrix and Hyper-T-matrix of a single molecule. This is to sufficiently incorporate the electronic interaction between the molecules in the TD-DFT simulations, which will modify the optical response as well. We continue to speak of it as the molecular unit cell. Our chosen model consists of 2x2x2 unit cells extracted from the crystallographic information framework (CIF) published by Guth \emph{et al.}\cite{GuthHegerKleinTreutmannScheringer+1980+237+254} as presented in Figure~\ref{fig:Urea_OPA}\textbf{(a)} and \textbf{(b)}. Upon converging the ground state electron density, the one-photon absorption (OPA) spectrum was constructed from calculated 200 lowest energy discrete transitions using the standard TD-DFT approach. The discrete electronic transitions (black vertical lines) together with its Lorentzian line-shape convoluted spectrum (violet) is visualized in Figure~\ref{fig:Urea_OPA}\textbf{(c)}. It demonstrates that Urea absorbs intensively only well below 300 nm in the UV part of the electro-magnetic spectrum. Furthermore, the two-photon absorption (TPA) spectrum was simulated (Figure~S1), which confirms almost non-existent cross-sections around 532 nm, the half-wavelength of commonly used lasers in nonlinear optical studies. This agrees with the known fact that Urea is a strong SHG material \cite{C8TC05268A,https://doi.org/10.1002/advs.202104379}. Additionally, we observe that it does not exhibit TPA for those wavelengths. 

To perform multi-scale simulations of the linear and, especially here, nonlinear response of the Urea material and a cavity device built thereof, we need to calculate the dynamic polarizability and hyperpolarizability tensors with TD-DFT. Figure~\ref{fig:Urea_OPA}\textbf{(d)} presents the simulated OPA cross-section of the selected parts of the electronic absorption based on the obtained polarizabilities. The spectral window for which the polarizabilities were calculated covers \textpm 100 nm around the excitation wavelength of the laser at 1064 nm, typically reported in Urea SHG experimental studies, as well as the spectral region which covers the second harmonic, i.e., 482-582 nm. In the infra-red part of the spectrum, the dynamic polarizabilities were calculated at a discretization of 2 nm. In the visible part, the discretization was 1 nm. The two parts of the spectrum (blue and red) in Figure~\ref{fig:Urea_OPA}\textbf{(d)} represent the same long-range tail of the Lorentzian line-shape of the high-energy electronic transitions below 300 nm. There are no resonances in these spectral windows.

\begin{figure*}
\centering
    
	\includegraphics[width=0.9\textwidth,keepaspectratio]{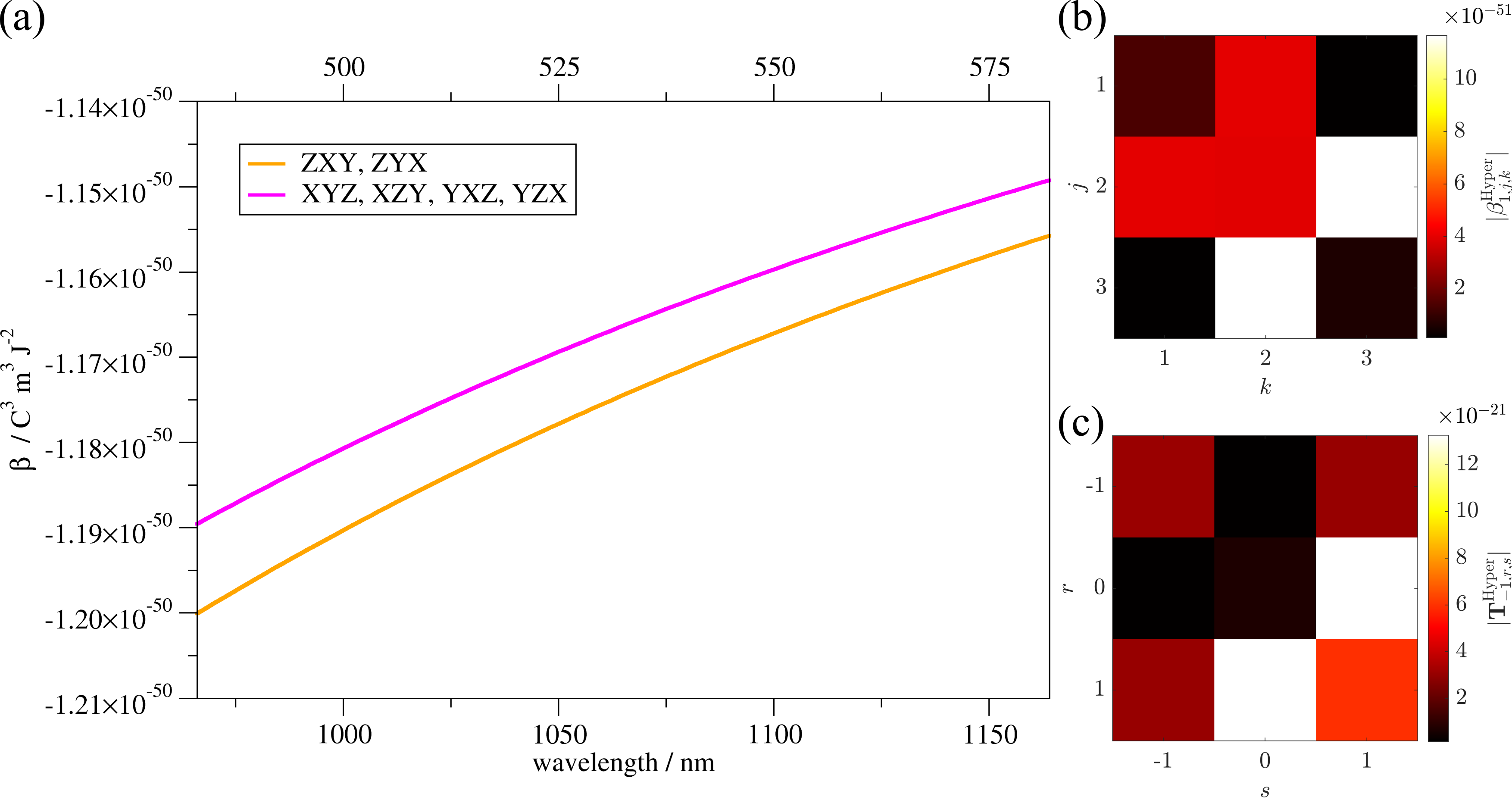}
 \hspace{0cm}

	\caption{\textbf{(a)} Real part of the first hyperpolarizabilities in SI units for the molecular model of Urea calculated by TD-DFT. \textbf{(b)} Compressed first hyperpolarizability tensor in the Cartesian basis and \textbf{(c)} Hyper-T-matrix in the spherical basis. Both matrices are highly anisotropic.}
    \label{fig:Color2D}
	\end{figure*}

We now present calculations of the electric-electric hyperpolarizabilities using the TD-DFT method. To perform those calculations and construct Hyper-T-matrices for the multi-scale approach, we extended the existing implementation of the calculation of the real first hyperpolarizabilities in the development version of the TURBOMOLE electronic structure package\cite{TURBOMOLE2022} to complex frequencies. Again, the same spectral range with the same wavelength step around 1064 nm was set. Figure~\ref{fig:Color2D}\textbf{(a)} plots the dominating components of the calculated first electric-electric hyperpolarizability tensor $\mathbf{\beta}$. In that spectral range of interest, the absolute value of $\mathbf{\beta}$ reduces only marginally with the increase in the wavelength. A graphical visualization of the absolute value of the elements of the hyperpolarizability tensor at 1064 nm is presented in Figure~\ref{fig:Color2D}\textbf{(b)} and of the Hyper-T-matrix in \textbf{(c)}, being compressed to the respective first entry. Note that the matrices are defined in different bases, namely the Cartesian and the spherical basis, respectively. We observe that both matrices are highly anisotropic. The presented approach takes this anisotropy entirely into account. After the successful construction of the T-matrices and Hyper-T-matrices of the Urea crystalline material from quantum chemistry calculations, we proceed with our multi-scale simulations of the thin film of Urea and of an optical cavity filled with the thin film of the same material.

\subsection{Thin film in vacuum}
In this subsection, we show the SHG response of a thin film of 100 layers of two-dimensional lattices consisting of a periodic arrangement of the unit cell of Urea. This corresponds to a molecular thin film with a thickness of 94\,nm. The response is compared to results obtained with COMSOL where the material is described using an effective permittivity and an effective second-order susceptibility. This serves the purpose of comparing our approach, which continues to treat the molecules as discrete nonlinear scatterer, to a traditional approach where the material is treated as a bulk medium characterized by macroscopic material parameters.

The considered thin film of the Urea molecular crystal is conceptionally shown in Figure~\ref{fig:Setup}\textbf{(a)}. 

\begin{figure*}
\centering
     \subfloat{
	\includegraphics[width=0.45\textwidth]{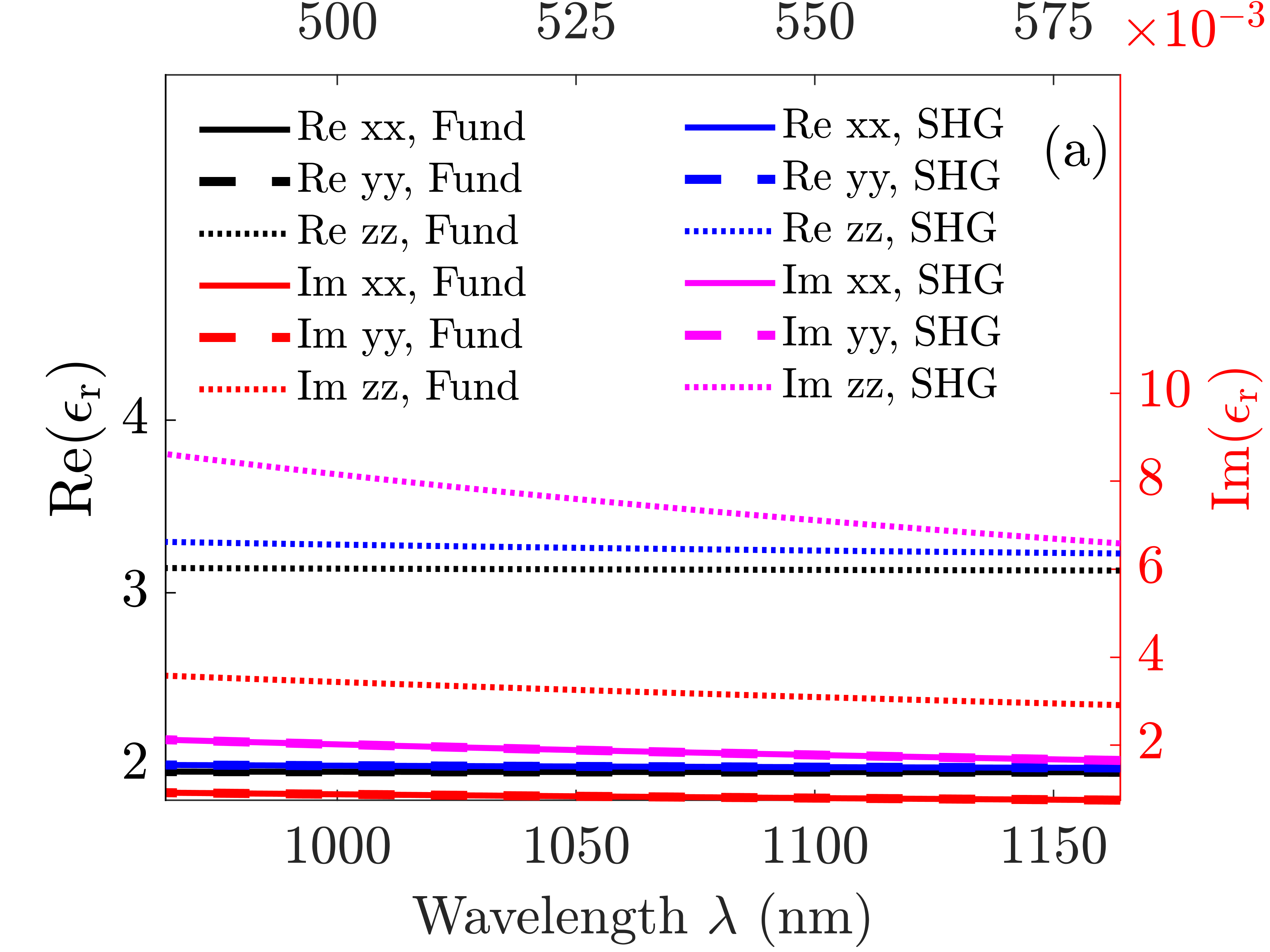}
 }\hspace{0cm}
 \subfloat{
	 \includegraphics[width=0.45\textwidth]{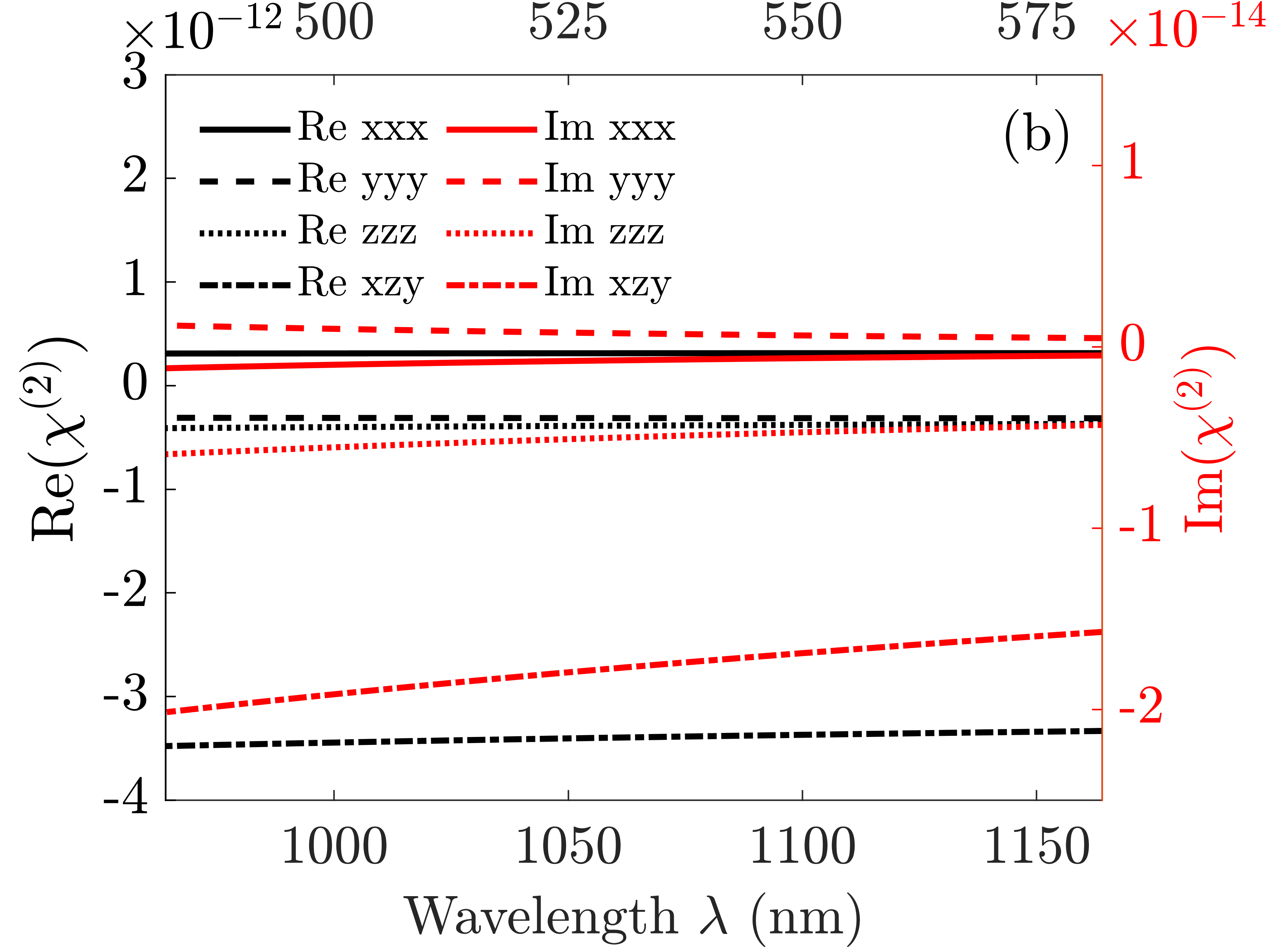}
}
	\caption{\textbf{(a)} Diagonal entries of the effective permittivity and \textbf{(b)} selected entries of the effective second-order susceptibility of the Urea molecular crystal. The lower x-axis refers to the wavelength of the incident fundamental fields, the upper x-axis refers to the wavelength of the SHG field. Both quantities are weakly dispersive and absorptive.}
    \label{fig:EffParam}
	\end{figure*}

For the effective medium considered in COMSOL, the effective material parameters are computed. For the calculation of the effective permittivity, the approach discussed in \cite{https://doi.org/10.1002/adom.202201564} is used. The second-order susceptibility is computed with Equation~(\ref{eq:chi}). The diagonal entries of the effective permittivity tensor are depicted in Figure~\ref{fig:EffParam}\textbf{(a)}. We observe that the material is characterized by weak dispersion and absorption. In Figure~\ref{fig:EffParam}\textbf{(b)}, selected entries of the effective second-order susceptibility are shown. Compared to experimental values published in literature \cite{C8TC05268A}, the relative difference in$|d_{14}^{\mathrm{Ref}}|=1/2\big|\mathbf{\chi}_{132}^{(2),\mathrm{Ref}}\big|=1.4\,\mathrm{pm}\,\mathrm{V}^{-1}$ is 21\,\% at the fundamental wavelength of 1064\,nm corresponding to a SHG wave at 532\,nm.

\begin{figure*}
\centering
     \subfloat{
	\includegraphics[width=0.45\textwidth]{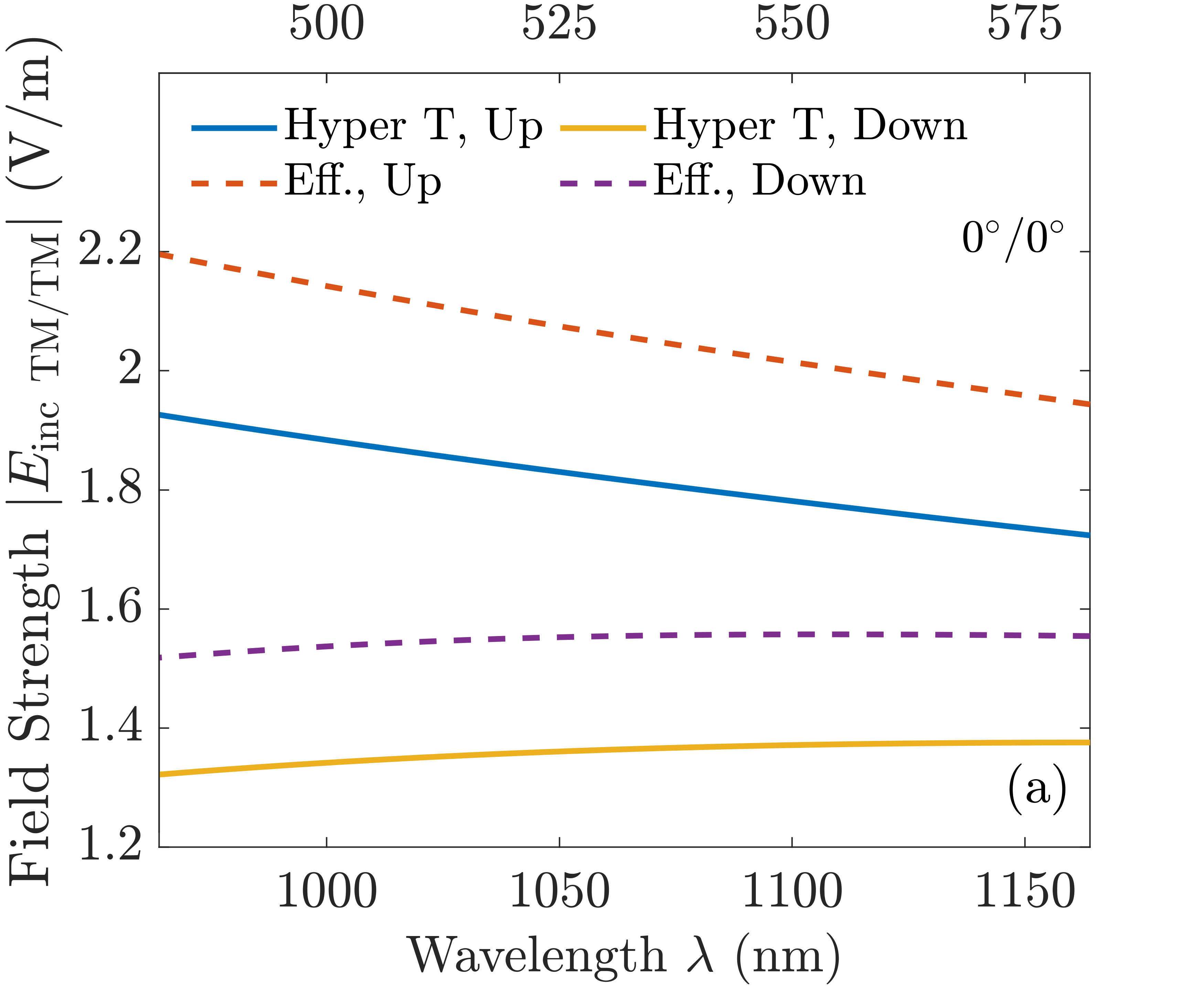}
 }\hspace{0cm}
 \subfloat{
	 \includegraphics[width=0.45\textwidth]{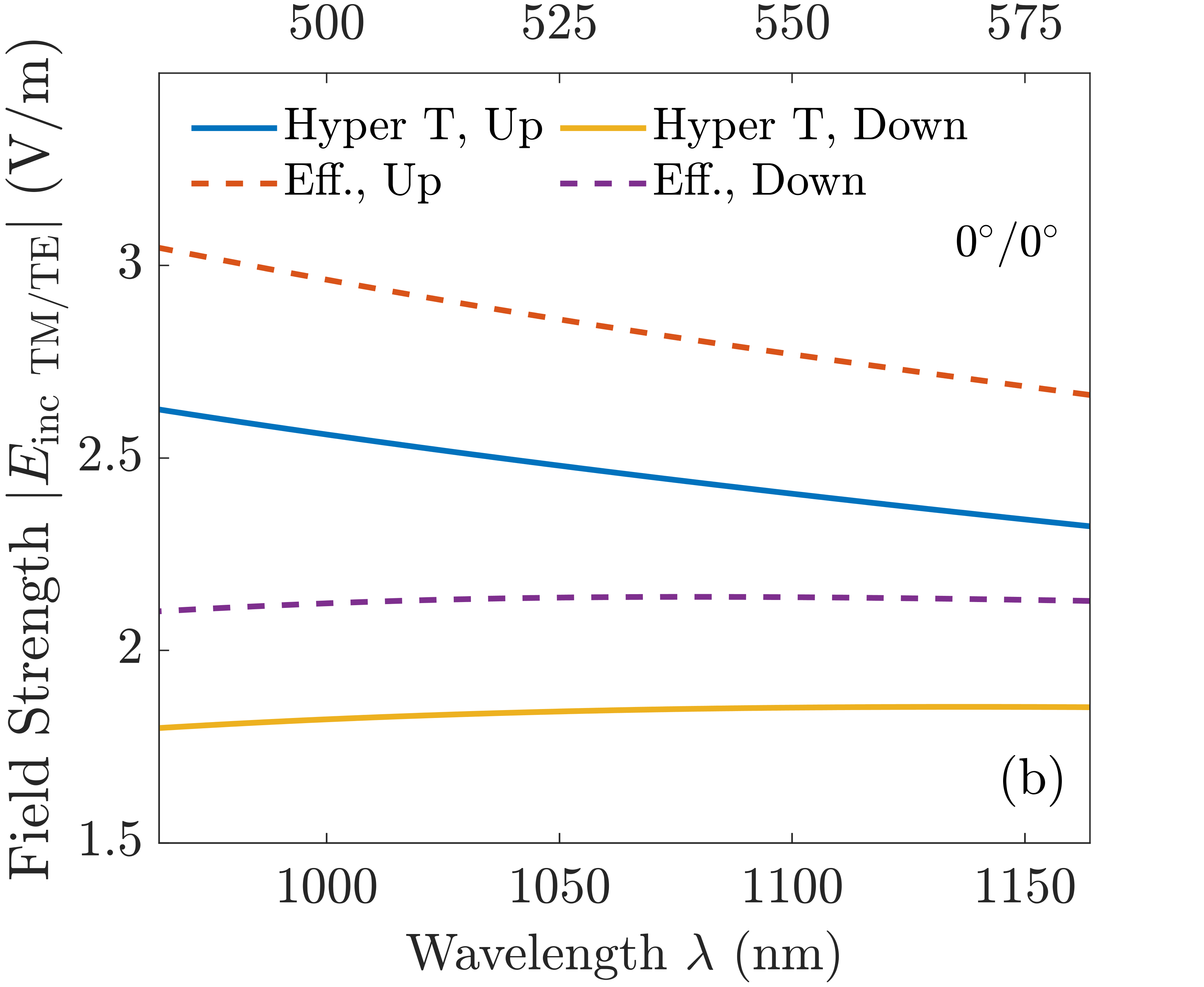}
}\\
\subfloat{
	\includegraphics[width=0.45\textwidth]{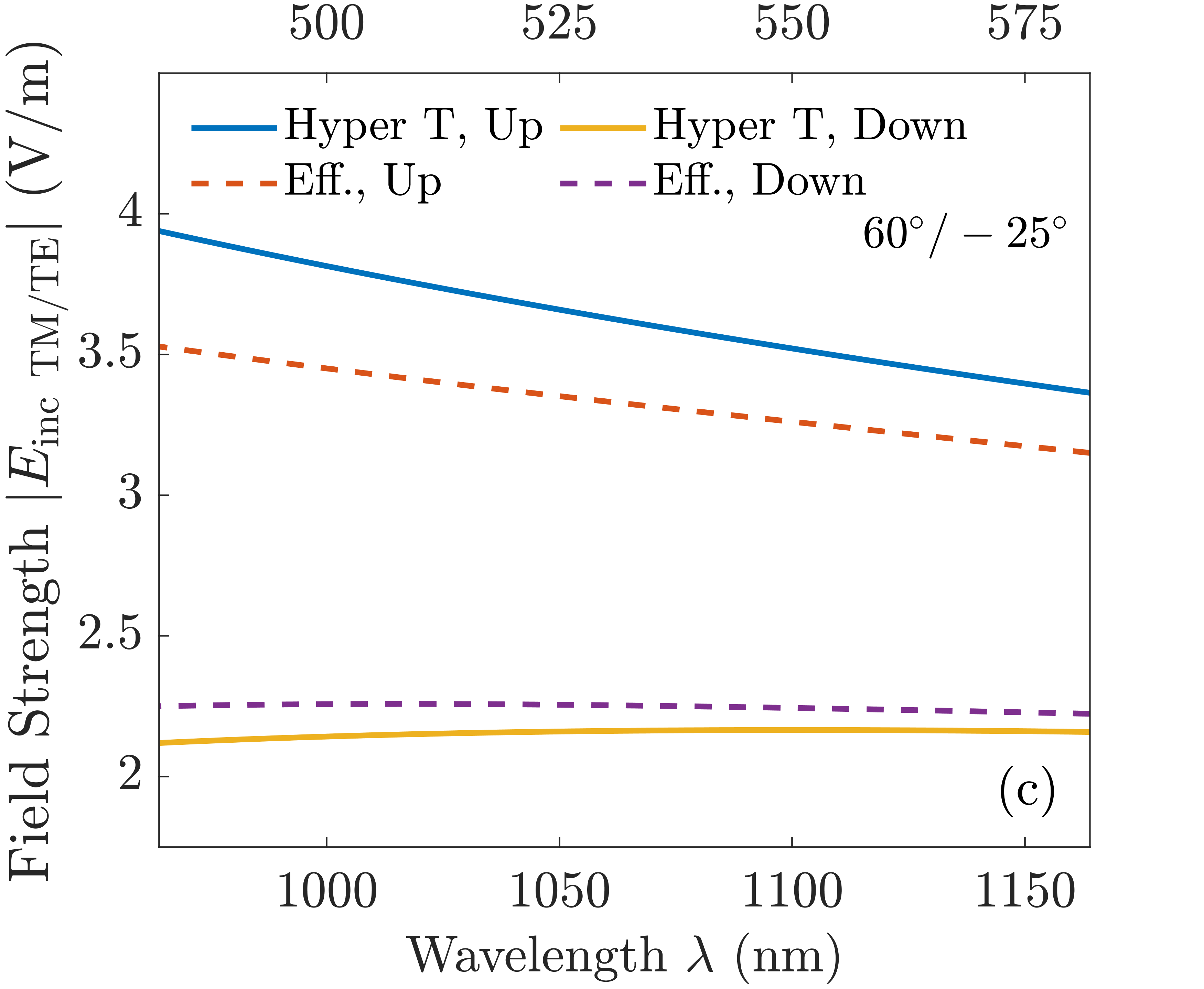}
 }\hspace{0cm}
 \subfloat{
	 \includegraphics[width=0.45\textwidth]{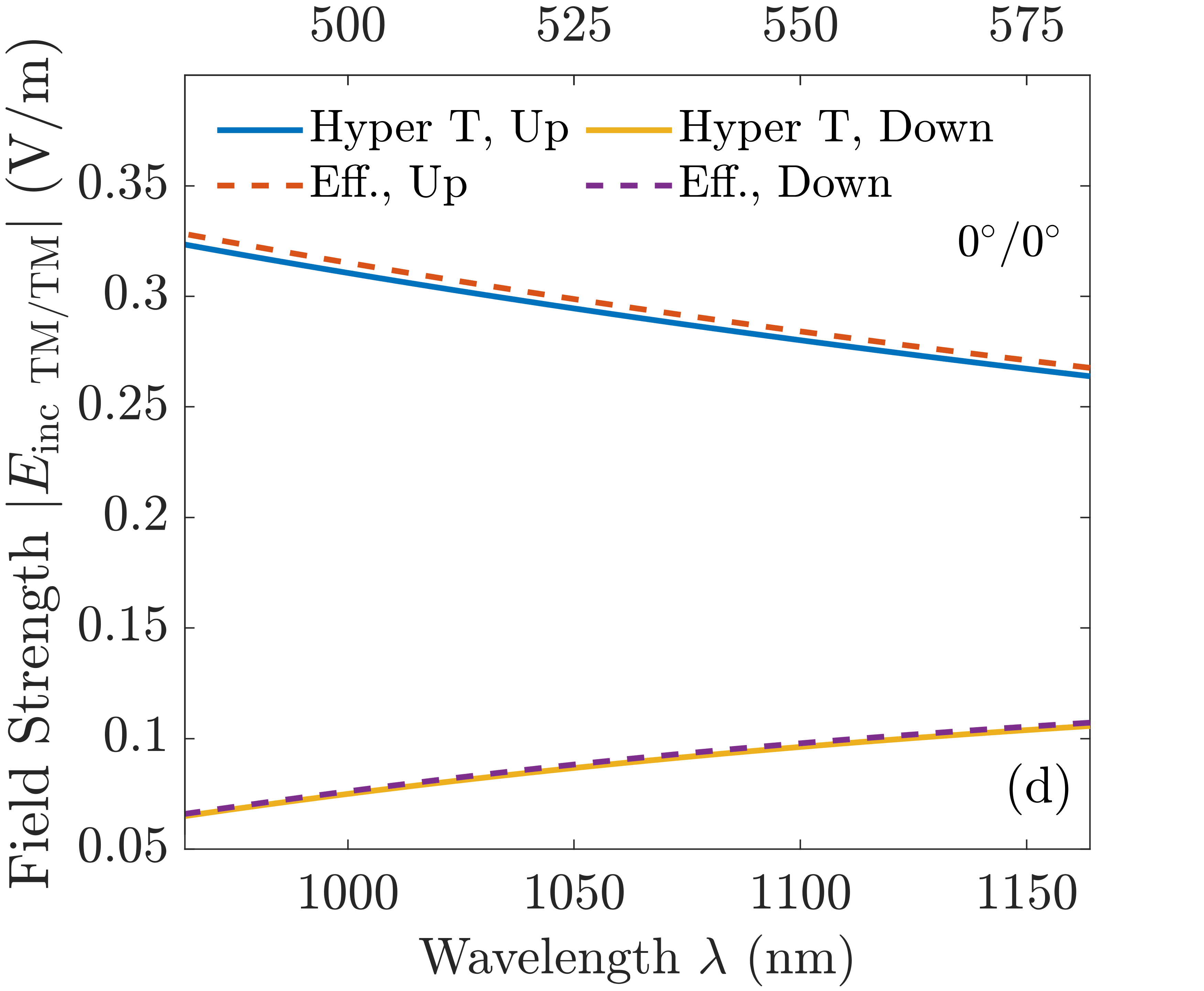}
}
	\caption{Amplitudes of the SHG plane wave generated at a film of 100 layers of a Urea molecular crystal in vacuum, propagating up- and downwards. The thin film has been illuminated by two TM-polarized plane waves at normal incidence in \textbf{(a)}, for a normal incident TM-polarized plane wave and a normal incident TE-polarized plane wave in \textbf{(b)}, for a TM-polarized plane wave with an angle of incidence of $60^{\circ}$ and a TE-polarized plane wave with an angle of incidence of $-25^{\circ}$ in the yz-plane in \textbf{(c)}. \textbf{(d)} shows the same up- and downward propagating plane waves when two normally incident TM-polarized plane waves excite a Urea molecular crystal thin film where the lattice constant was artificially doubled. In all figures, the results for the presented Hyper-T-matrix based approach (solid lines) and for the finite-element based method using an effective second-order susceptibility tensor are compared (dashed lines). In \textbf{(a)}-\textbf{(c)}, the maximum difference is 16.86\,\%. The approach where the material is described at the level of effective bulk material parameter neglects the optical interaction between the molecular unit cells of the crystal. Additionally, the smallest circumscribing spheres of adjacent unit cells overlap decreasing the accuracy of the T-matrix formalism. When increasing the lattice constants, both the interaction and the overlap of adjacent unit cells reduce, which causes the excellent agreement of both descriptions in \textbf{(d)}.}
    \label{fig:CompEff}
	\end{figure*}

In Figure~\ref{fig:CompEff}\textbf{(a)}-\textbf{(c)}, the amplitudes of the up- (in positive z-direction) and downwards (in negative z-direction) propagating SHG plane waves are shown for both incident plane waves being normally incident and TM polarized \textbf{(a)}, one being TM- and the other one being TE-polarized \textbf{(b)}, and one being TM-polarized having an angle of incidence of $60^{\circ}$ and one being TE-polarized having an angle of incidence of $-25^{\circ}$ in the yz-plane \textbf{(c)}, respectively. At normal incidence, TM-polarization corresponds to a polarization along the x-axis, and TE-polarization corresponds to a polarization along the y-axis. For oblique incidence, the fields are rotated accordingly. 

We observe a fairly good match between the results obtained with the presented approach based on the Hyper-T-matrix (solid lines) and with the finite-element-based computation using the effective second-order susceptibility tensor (dashed line). The maximum relative difference is 16.86\,\%. Discrepancies between the results can be related to two aspects. First, Equation~(\ref{eq:chi}) for the second-order susceptibility used in COMSOL neglects the optical interaction between different unit cells. Second, the smallest circumscribing sphere of neighboring unit cells overlaps to some extent, particularly in the z-direction. As the T-matrix-based computation of the interaction between different scatterers assumes that those so-called Rayleigh spheres do not overlap \cite{Peterson1973,MISHCHENKO1996,Schebarchov:19}, the interaction between the molecular unit cells cannot be computed as accurate as for non-penetrating circumscribing spheres. This clearly indicates some advantages and disadvantages of each of the methods. 

For the spectra in Figure~\ref{fig:CompEff}\textbf{(d)}, we artificially double the lattice constants. This decreases (a) the optical interaction between the unit cells and (b) their Rayleigh spheres do not overlap anymore. In this case, the match between both results is excellent, clearly indicating the consistency between both approaches. 

With these results, we demonstrate that the approach based on the Hyper-T-matrix can be used to compute the SHG response of a molecular film for different polarizations and angles of incidence of illuminating plane waves. This is especially important if a material is anisotropic and highly dispersive. A second aspect is that we do not only compute the response for a specific wavelength of a specific excitation laser but for an extended spectral domain. This enables the computation of the SHG response in wavelength ranges that are not accessible by nowadays lasers but possibly by future light sources.

\subsection{Optical cavity filled with thin film}
\begin{figure*}
\centering
     \subfloat{
	\includegraphics[width=0.45\textwidth]{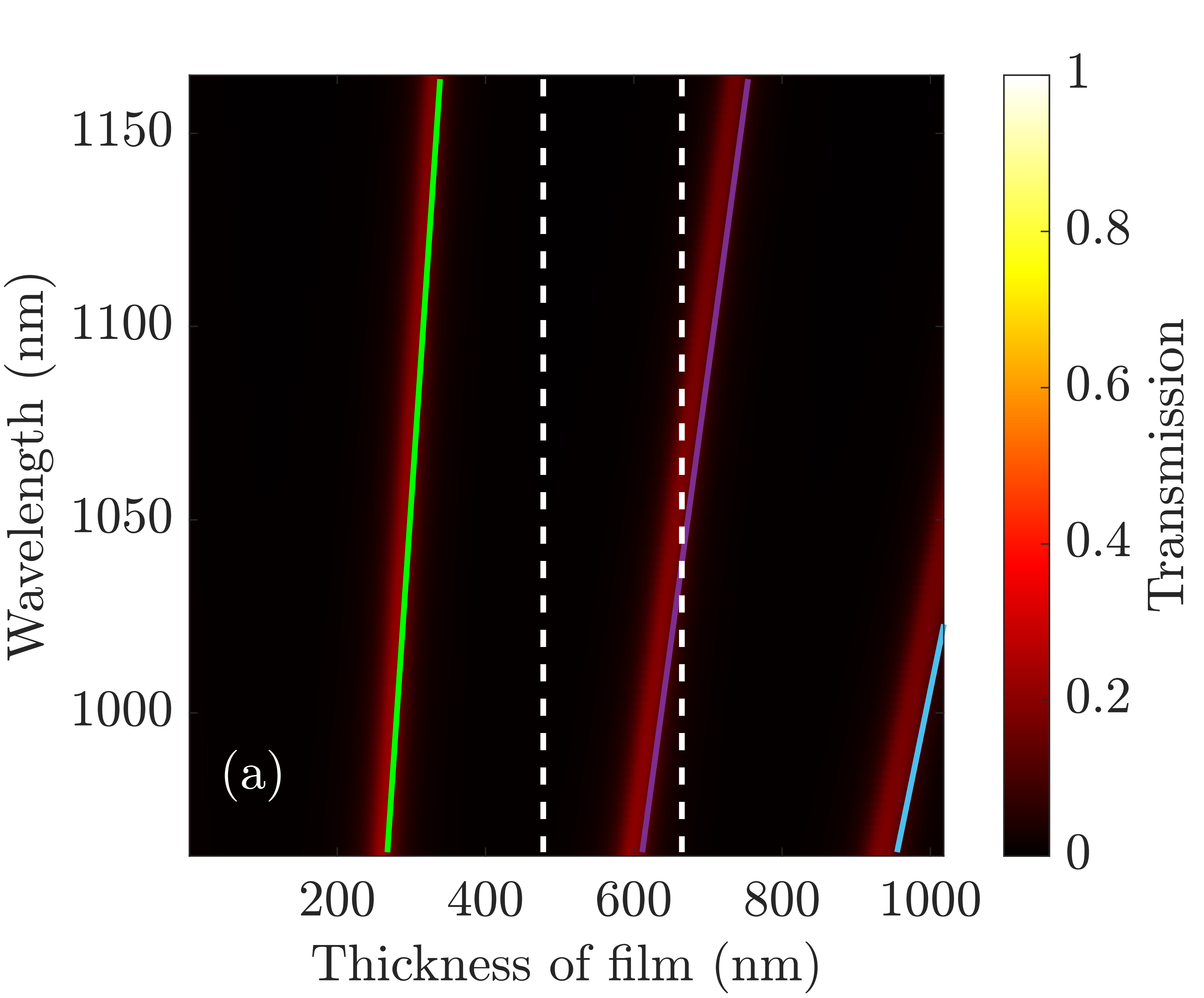}
 }\hspace{0cm}
 \subfloat{
	\includegraphics[width=0.45\textwidth]{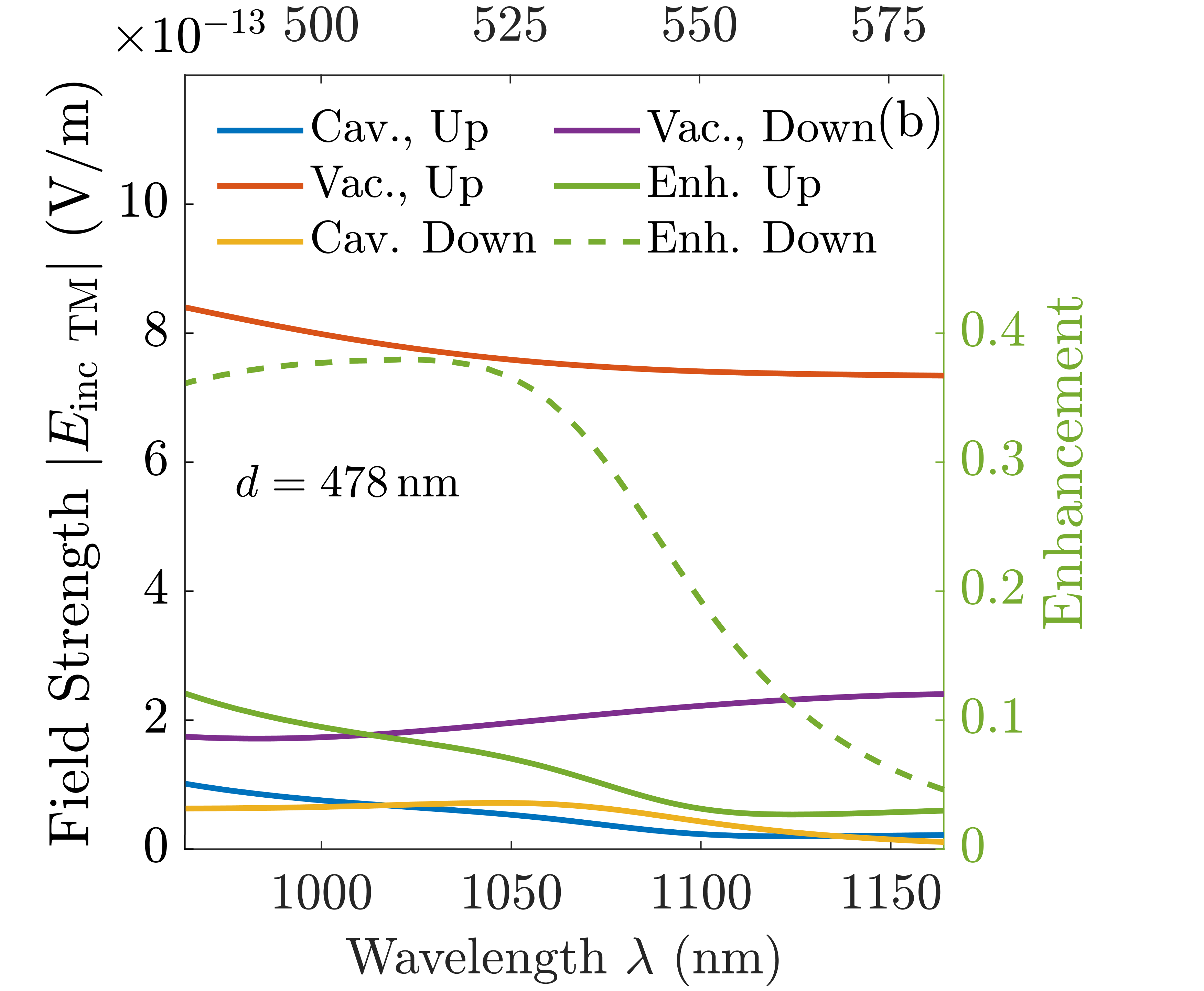}
 }
 \\
 \subfloat{
	 \includegraphics[width=0.45\textwidth]{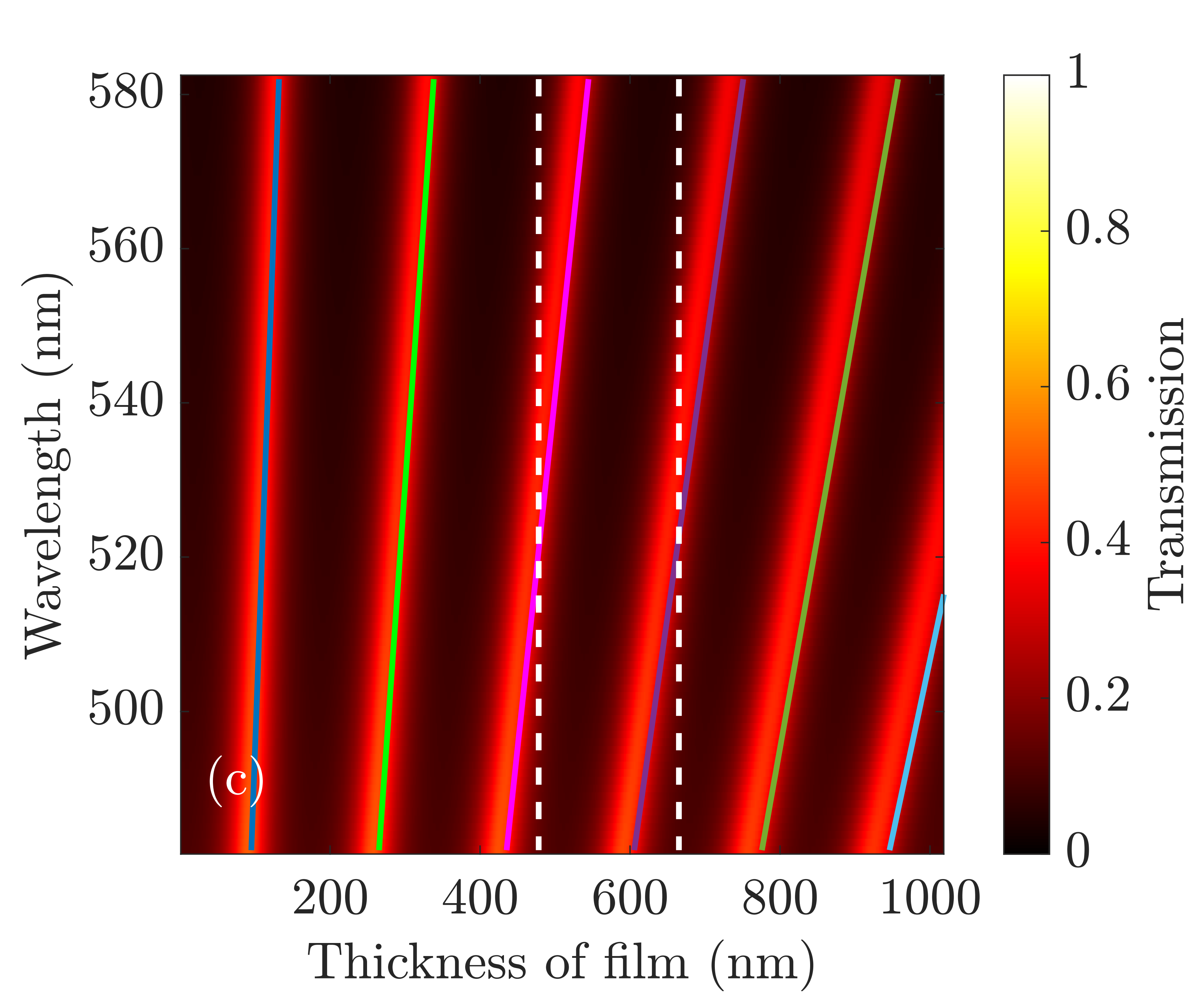}
}
\hspace{0cm}
 \subfloat{
	 \includegraphics[width=0.45\textwidth]{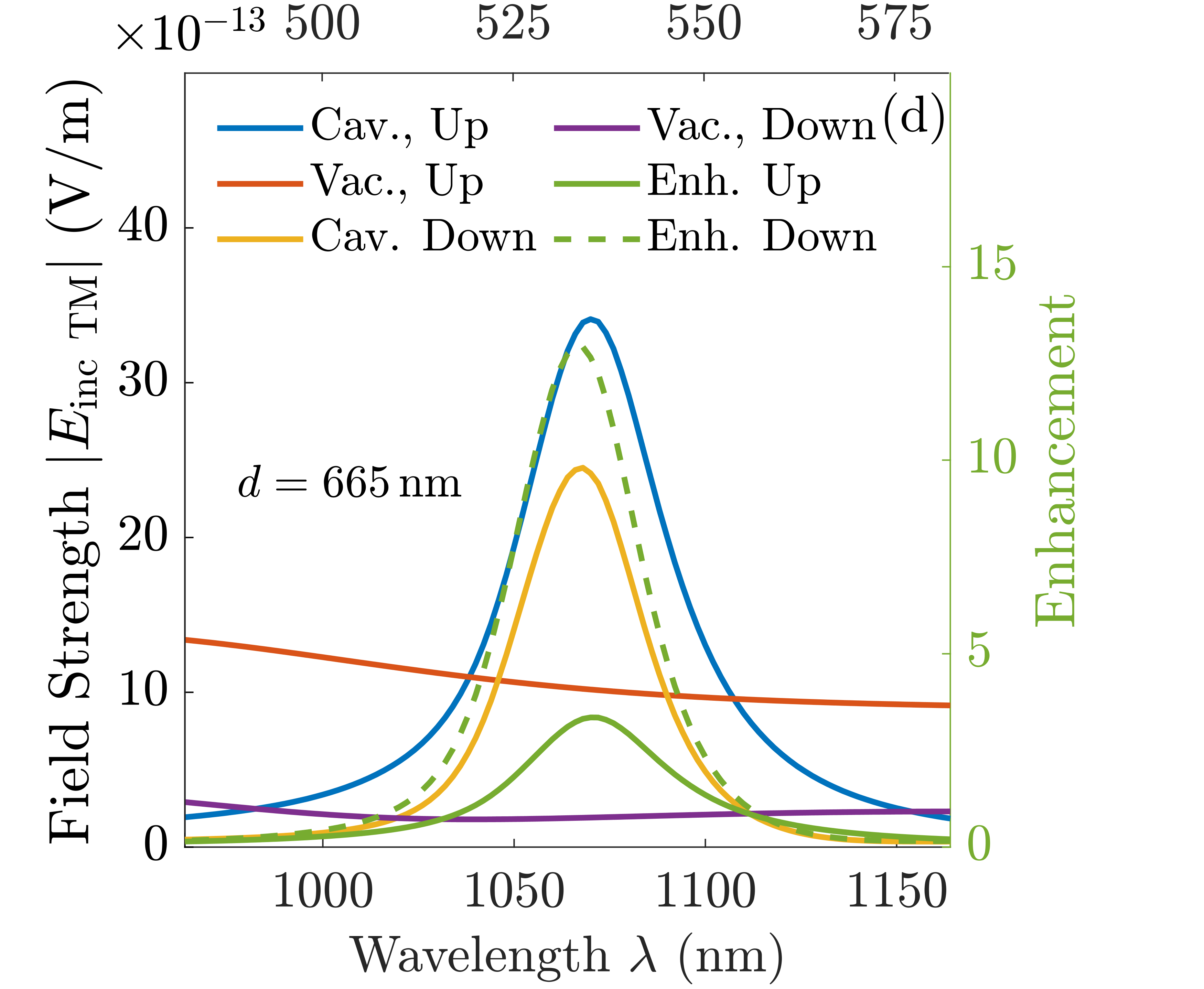}
}
	\caption{Linear transmission of an optical cavity filled with a film of the Urea molecular crystal for a normal incident TM-polarized plane wave with the fundamental frequency \textbf{(a)} and the same plane wave with the SHG frequency \textbf{(c)}. Amplitudes of the SHG plane waves of the filled cavity propagating up- and downwards for a thickness of the film of 478\,nm \textbf{(b)} and 665\,nm \textbf{(d)}. The thicknesses correspond to the white lines in \textbf{(a)} and \textbf{(c)}. For 478\,nm, the cavity sustains only a mode at the SHG frequency. For 665\,nm the cavity sustains modes at the fundamental and the SHG field. Clearly, the SHG signal is enhanced in the latter case.}
    \label{fig:Cavity}
	\end{figure*}
To demonstrate further application aspects of the presented approach, we consider the optical cavity shown in Figure~\ref{fig:Setup}\textbf{(b)}. It consists of two silver mirrors separated by a thin film made from the Urea molecular crystal with variable thickness. The lower mirror has a thickness of 10\,nm, the upper mirror has a thickness of 30\,nm. The optical parameters of the mirrors are taken from \cite{Jiang:2016}. We aim to enhance the SHG signal of the molecular film for a specific design wavelength by cavity modes. The enhancement of SHG signals and other nonlinear optical signals of various samples by optical cavities has been considered both theoretically and experimentally in several studies \cite{doi:10.1021/acs.nanolett.1c03824,doi:10.1021/acsnano.2c03033,PhysRevA.99.043844,PhysRevA.104.063502,PhysRevLett.127.153901,Barclay:05,McLaughlin:22}, as the nonlinear optical signal of natural materials is usually rather small and difficult to detect. 

To understand the modal structure, the linear transmission is initially computed for an illuminating TM-polarized plane wave at the fundamental and the SHG frequencies, respectively, and for varying thicknesses of the molecular film. In Figure~\ref{fig:Cavity}\textbf{(a)} and \textbf{(c)}, both transmission spectra are shown with the cavity modes computed with an analytical mode analysis described in \cite{SURMOFCavity}. For the latter, the refractive index of the x-axis is used, neglecting the anisotropy of the film. We observe that every cavity mode at the fundamental frequencies coincides with a cavity mode at the SHG frequencies. This is because of the marginal dispersion of the molecular material. At the SHG frequencies, twice the number of cavity modes are present compared to the fundamental frequencies. That behavior reflects the fact that in the SHG process, the frequency of the nonlinear field is twice the frequency of the fundamental field.

In the second step, the amplitudes of the up- and downward propagating SHG plane waves are computed for specific thicknesses of the molecular thin film. We choose for the thickness 478\,nm (corresponding to 509 layers) and 665\,nm (708 layers). For the first thickness, the third cavity mode at the SHG wavelength 532\,nm is supported, see Figure~\ref{fig:Cavity}\textbf{(c)}. In this case, the cavity does not support a cavity mode at the fundamental wavelength of 1064\,nm. For the latter thickness of 665\,nm, the cavity supports its second-order cavity mode at the fundamental frequency, see Figure~\ref{fig:Cavity}\textbf{(a)}, and a higher-order cavity mode at the SHG frequency, i.e., it is double resonant. 

In Figure~\ref{fig:Cavity}\textbf{(b)},\textbf{(d)}, the SHG amplitudes and their enhancement compared to the thin film in a vacuum are shown. We observe that only in the double resonant scenario, the cavity enhances the SHG fields with maxima close to the design wavelength of 1064\,nm. This is because in the case of the single resonant cavity in Figure~\ref{fig:Cavity}\textbf{(b)}, the field of the illuminating fundamental wave is decreased compared to the case of a thin film in vacuum due to destructive interference. In the double resonant scenario, the fields at the fundamental and at the SHG frequency are enhanced inside the molecular thin-film, which enhances the generated SHG amplitude by 3.11 in the upwards and 12.75 in the downwards direction at the design wavelength of 1064\,nm of the illuminating fundamental wave. 

In this subsection, we have shown that our approach can be used to design an optical cavity consisting of metal mirrors to enhance the SHG response of the molecular material. This is very important, as the SHG signal of natural materials is usually very difficult to detect.

\section{Conclusions and Outlook}\label{sec:Conclusion}
We have presented a novel multi-scale approach to compute the nonlinear response of molecular nanomaterials in photonic devices. For this purpose, we introduced an object called the Hyper-T-matrix, which can be computed from the first hyperpolarizability. The first hyperpolarizability of a molecular unit cell is calculated with the \textit{ab initio} method TD-DFT. We demonstrated the approach by studying a thin-film made from the Urea molecular crystal in a vacuum and upon integration into an optical cavity. An enhancement of the intensity of the SHG signal of 167 was achieved by specifically designing the cavity. This demonstrates the usefulness of our computer-aided design. Nonlinear optical signals are usually very small, and it is decisive to enhance them for practical purposes. Now, one can simulate the response of a novel molecular material or photonic device already beforehand instead of performing demanding experiments.

Future research can focus on other second-order nonlinear processes, such as sum-frequency generation, for instance, on higher-order nonlinear processes, such as third-order harmonic generation, or on the design of new molecular materials or photonic devices with a tailor-made optical response. Since our approach is general, it can be straightforwardly extended toward other nonlinear processes.

Moreover, with the availability to express the nonlinear optical response of a single molecule in terms of the Hyper-T-matrix, molecules could be considered in more advanced nanophotonic systems. For example, studying the nonlinear response of individual molecules in nanophotonic cavities made from, e.g., closely touching spheres or nanoparticles-on-substrates, could be a fascinating research question. Also, it would be interesting to probe the emergence of a second-order nonlinear response close to interfaces from molecules that are centro-symmetric in the bulk. All these questions can be tackled with our novel multi-scale modeling ansatz to treat nonlinear molecular materials.

\medskip
\textbf{Acknowledgements} \par 
M.K., D.B., and C.R. acknowledge support by the Deutsche Forschungsgemeinschaft (DFG, German Research Foundation) under Germany’s Excellence Strategy via the Excellence Cluster 3D Matter Made to Order (EXC-2082/1-390761711) and from the Carl Zeiss Foundation via the CZF-Focus@HEiKA Program. M.K., C.H., and C.R. acknowledge funding by the Volkswagen Foundation. I.F.C. and C.R. acknowledge support by the Helmholtz Association via the
Helmholtz program “Materials Systems Engineering” (MSE). B.Z. and C.R. acknowledge support by the KIT through the “Virtual Materials Design” (VIRTMAT) project. M.K. and C.R. acknowledge support by the state of Baden-Württemberg through bwHPC and the German Research Foundation (DFG) through grant no. INST 40/575-1 FUGG (JUSTUS 2 cluster) and the HoreKa supercomputer funded by the Ministry of
Science, Research and the Arts Baden-Württemberg and by the Federal Ministry of Education and Research. 

\medskip
\textbf{Conflict of Interest}
The authors declare no conflict of interest.

\medskip
\textbf{Data Availability Statement}
The data that support the findings of this study are available from the
corresponding authors upon reasonable request. The complete set of quantum chemistry calculations based on TD-DFT method produced for this study is deposited in the NOMAD materials science database under the following DOI link: \href{https://doi.org/10.17172/NOMAD/2023.08.31-1}{https://doi.org/10.17172/NOMAD/2023.08.31-1} 

\bibliography{bibliographyArxiv}

\end{document}